\documentstyle[graphicx,amsmath,amssymb,fleqn]{mn}

\newcommand{\txd}{\text{d}}
\renewcommand{\leq}{\leqslant}
\renewcommand{\geq}{\geqslant}
\DeclareMathOperator{\sech}{sech}
\title[Radiative transfer in disc galaxies II]{Radiative transfer in
disc galaxies II -- The influence of scattering and geometry on the
attenuation curve}

\author[Baes \& Dejonghe]{Maarten Baes\thanks{Research Assistent of
the Fund of Scientific Research -- Flanders (Belgium)} and Herwig
Dejonghe \\ Sterrenkundig Observatorium Universiteit Gent, Krijgslaan
281 S9, B-9000 Gent, Belgium, maarten.baes@rug.ac.be }

\begin{document}

\maketitle
\begin{abstract}
We investigate the influence of scattering and geometry on the
attenuation curve in disc galaxies. We investigate both qualitatively
and quantitatively which errors are made by either neglecting or
approximating scattering, and which uncertainties are introduced due
to a simplification of the star-dust geometry. We find that the
magnitude of these errors depends on the inclination of the galaxy, and
in particular that for face-on galaxies, the errors due to an improper
treatment of scattering dominate those due to an imprecise star-dust
geometry. Therefore we argue that in all methods which aim at
determining the opacity of disc galaxies, scattering should be taken
into account in a proper way.
\end{abstract}


\section{Introduction}

During the last decade it has become clear that also in extragalactic
astronomy, dust obscuration plays an important role, and that the
construction of accurate galaxy models requires a proper radiative
transfer modelling. However, it is often very time-consuming, and
therefore practically unfeasible, to solve radiative transfer problems
exactly. For example, assume we want to know the spatial distribution
of stars and dust for a certain galaxy, where the only known data are
observed surface brightnesses. The only means to solve this problem is
to assume a certain parametrized model for the galaxy, containing a
stellar and a dust component, and to solve the radiative transfer
equation (RTE) for each set of parameters, and then select the model
that fits the observations best. Such a modelling procedure demands an
enormous number of RTE solutions, because the parameter space has to
be sufficiently large in order to produce realistic models, in
particular for disc galaxies (Misiriotis et al.\ 2000). Therefore it
would be very useful if we could simplify the RTE in any way.

A first way to simplify the RTE is rather obvious: neglect the
scattering term, which is responsible for the integro-differential
character of the RTE (Christensen 1990, Jansen et al.\ 1994, Ohta \&
Kodaira 1995). However, already more than a decade ago, Bruzual et
al.\ (1988) clearly demonstrate the importantance of including
scattering in radiative transfer calculations. Several other authors
have since then confirmed these warnings, in particular Witt et al.\
(1992). Another option to simplify the RTE is an approximation of the
scattering term, by either forward or approximate isotropic
scattering, or a combination of both (Natta \& Panagia~1984,
Guiderdoni \& Rocca-Volmerange~1987, Calzetti et al.\ 1994). Di
Bartolomeo et al.\ (1995) compare the attenuation curves calculated
with these various approximate RTE solution methods, and find
significant differences. Although these differences can partly be
ascribed to different assumptions concerning the optical properties of
the dust, it is very probable that also the approximate solution
methods are responsible for a significant error.

It is obviously important to know the errors introduced on the
attenuation curve by the various approximate solution techniques, but
these have never been clearly determined. Either the approximate
solutions are simply adopted without concern about the errors, or the
errors are estimated for very simple galaxy models, such as
homogeneous slabs or sandwich models (Bruzual et al.\ 1988, Di
Bartolomeo et al.\ 1995). To make things worse, Disney et al.\ (1989)
showed that the geometry of the stellar and dust distribution strongly
determines the observed radiation field (without including the effects
of scattering however). Lack of knowledge about the geometry can thus
also introduce significant errors on the attenuation of galaxies.

In a previous paper (Baes \& Dejonghe~2001, paper~I) we described four
different methods to solve the RTE in a plane-parallel geometry. These
methods accomodate an arbitrary vertical distribution of stars and
dust. We will now use these methods will allow us to address both
issues of the RTE mentioned above. On the one hand, we can adopt them
for a disc galaxy model with a realistic vertical structure, and
compare the results with those obtained by using the various
approximate solutions. This will allow us to quantify the errors
introduced by the different approximations, and the importance of
properly including scattering effects. On the other hand, we can apply
these methods to a {\em family} of realistic galaxy models that can
accomodate a wide range in distribution of stars and dust. This will
allow us to investigate the influence of the geometry of the stellar
and dust components on the attenuation curve, without any simplifying
assumptions on the RTE such as the neglect of scattering. Last but not
least, we can solve our RTE problems with four completely different
methods: consistency is then a guarantee for accuracy.

In Section~2 we describe the radiative transfer mechanism and the ways
to obtain the solution, and in Section~3, we present our set of disc
galaxy models. In Section~4 and~5 we discuss respectively the
influence of scattering and geometry on the attenuation curve as
described above. In Section~6 we discuss the results.

\section{Radiative transfer models}

\subsection{The RTE in plane-parallel geometry}

In plane-parallel geometry, the RTE can be written as
\begin{multline}
	\mu\,\frac{\partial I}{\partial z}(z,\mu)
	=
	-\kappa(z)I(z,\mu)
	+\eta_*(z) 
	\\ 
	+\frac{1}{2}\omega\kappa(z)
	\int_{-1}^1I(z,\mu')\Psi(\mu,\mu')\txd\mu',
\label{tv1}
\end{multline}
where $I(z,\mu)$ is the specific intensity of the radiation at a
height $z$ above the plane of the galaxy, and in a direction which
makes an angle $\arccos\mu$ with the face-on direction $\mu=1$. The
known quantities in this equation are the dust opacity $\kappa(z)$,
the stellar emissivity $\eta_*(z)$, the dust albedo $\omega$ and the
angular redistribution function (hereafter ARF) $\Psi(\mu,\mu')$.  The
RTE can be brought in another form by introducing the optical depth
$\tau$ instead of $z$,
\begin{equation}
	\tau(z) 
	= 
	\int_z^\infty \kappa(z')\txd z',
\label{optdepth}
\end{equation}
which yields
\begin{equation}
	\mu\,\frac{\partial I}{\partial\tau}(\tau,\mu)
	=
	I(\tau,\mu)
	-S_*(\tau) 
	-\frac{\omega}{2}
	\int_{-1}^1I(\tau,\mu')\Psi(\mu,\mu')\txd\mu',
\label{tv2}
\end{equation}
with $S_*(\tau) = \eta_*(\tau)/\kappa(\tau)$
is the stellar source function. This equation is to be solved for
$0\leq\tau\leq\tau_0$, the total optical (face-on) depth of
the galaxy,
\begin{equation}
	\tau_0
	= 
	\int_{-\infty}^\infty \kappa(z)\txd z.
\label{deftau0}
\end{equation}
Although the RTE can be solved for any point in the galaxy, we will
focus on the attenuation $A(\mu)$, i.e.\ the fraction of the intensity
that is attenuated by the dust detected by an observer at $\tau=0$,
into a certain direction $\mu$. Because the RTE is a linear equation,
this fraction will be independent of the total amount of stellar
emission. We can thus choose the normalization of the stellar
emissivity (or the source function). We take
\begin{equation}
	\int_{-\infty}^{\infty}\eta_*(z)\,\txd z
	=
	\int_0^{\tau_0}S_*(\tau)\,\txd\tau
	=
	1,
\label{normeta}
\end{equation}
which means that the intensity that leaves the galaxy in the absence
of dust in the face-on direction equals 1. In another direction $\mu$,
the dust-free intensity is then simply $1/\mu$ and the attenuation (in
magnitudes) is
\begin{equation}
	A(\mu) 
	= 
	-2.5\log\left[\mu\,I(0,\mu)\right].
\label{defA}
\end{equation}

\subsection{Solution of the RTE}

Because of the complexity of the RTE, a lot of authors have tried
their ingenuity to find sophisticated methods to solve it. In paper~I
we presented four different methods to solve the RTE in plane-parallel
geometry, which can handle absorption and multiple scattering,
arbitrary vertical distributions of stars and dust and arbitrary phase
functions.
\begin{enumerate}
\item The spherical harmonics method consists of expanding all the
angle-dependent terms in the RTE into a series of spherical harmonics,
such that the RTE is replaced by a set of ordinary differential
equations. The method turns out to be very efficient.
\item In the discretization method, originating from the theory of
stellar atmospheres, integrals are replaced by sums and differentials
by finite differences, resulting in a set of vector equations which
can be solved iteratively.
\item In the iteration method, the intensity is expanded in a series
of partial intensities. Each of the partial intensities obeys its own
RTE, which can be solved iteratively. It can easily be extended to
more complex geometries, but compared to the spherical harmonics
method it is a very costly algorithm.
\item The Monte Carlo method is probably the most widely adopted
method to solve radiative transfer problems, and the basic idea of the
method consists of following the trajectory of a large amount of
individual photons through the galaxy, whereby the fate of a photon on
its path is determined by random events. For one-dimensional radiative
transfer its efficiency is comparable to that of the iteration method.
\end{enumerate}
For more details about these methods and references to the literature
we refer to paper~I. In this paper we will study the radiative
transfer through a plane-parallel slab, as a function of the different
physical and geometrical parameters.  Having these four different
methods to solve the RTE at our disposal proves to be very useful, not
only to check the accuracy of the numerical results, as already
mentioned, but also to understand the physical background of observed
phenomena.

\section{The disc galaxy models}

In order to construct disc galaxy models we have to characterize the
functions that appear in the RTE~(1), i.e.\ specify the physical
properties and spatial distribution of stars and dust. We will adopt a
set of realistic galaxy models with a number of parameters. Varying
these parameters will then enable us to investigate the influence of
scattering and geometry on the attenuation curve. We will also adopt a
template model to compare other models with.

\subsection{The optical properties of the dust}


\begin{table}
\centering
\caption{The adopted data sets for the optical properties of the dust
grains. Tabulated are the optical depth $\tau$ relative to the
$V$-band value, the scattering albedo $\omega$ and the asymmetry
parameter $g$.}
\label{optprop.tab}
\begin{tabular}{ccccc} \hline 
band & $\lambda$\,($\mu$m) & $\tau$ & $\omega$ & $g$ \\ \hline
UV1 & 125 & 3.44 & 0.42 & 0.60 \\
UV2 & 150 & 2.75 & 0.42 & 0.59 \\
UV3 & 175 & 2.44 & 0.48 & 0.58 \\
UV4 & 200 & 2.87 & 0.48 & 0.54 \\
UV5 & 225 & 3.01 & 0.51 & 0.46 \\
UV6 & 250 & 2.38 & 0.57 & 0.46 \\
UV7 & 300 & 1.92 & 0.57 & 0.47 \\
$U$ & 360 & 1.60 & 0.57 & 0.49 \\
$B$ & 440 & 1.32 & 0.57 & 0.48 \\
$V$ & 550 & 1.00 & 0.57 & 0.44 \\
$R$ & 700 & 0.73 & 0.54 & 0.37 \\
$I$ & 850 & 0.47 & 0.51 & 0.31 \\ \hline 
\end{tabular}
\end{table}


Although it has been shown that the physical properties of dust in
different environments can vary greatly (Witt et al.\ 1984, Mathis \&
Cardelli 1992), we will assume, for sake of simplicity, one single
kind of dust grains. This means that the spatial and wavelength
dependencies of all quantities appearing in~(1) are separable. In
particular, the dust albedo and the ARF are then independent of
position.

To describe general anisotropic (conservative) scattering we adopt
Henyey-Greenstein scattering (Henyey \& Greenstein~1941). The ARF
corresponding to this kind of scattering is a one-parameter function
parametrized by the asymmetry parameter $g$, which is the average of
the cosine of the scattering angle. A closed expression and a plot of
the Henyey-Greenstein ARF can be found in Appendix~A of Paper~I. In
Section~{\ref{scatimpo.sec}} other kinds of scatterings such as
forward and isotropic scattering, will be compared with
Henyey-Greenstein scattering.

There are two ways to determine the wavelength dependence of the
albedo $\omega$, the asymmetry parameter $g$ and the total
optical depth $\tau_0$. On the one hand, values can be derived
theoretically, by assuming a certain dust grain composition and
calculating the optical properties of the dust (e.g.\ Draine \&
Lee~1984). On the other hand, optical properties can be derived
empirically, usually based on a variety of observations of scattered
light in the Galaxy (e.g.\ Bruzual et al.\ 1988). Our data set
consists of the optical properties theoretically derived by Maccioni
\& Perinotto (1994), displayed in Di Bartolomeo et al.\ (1995). The
adopted values are tabulated in table~{\ref{optprop.tab}} for the
central wavelengths of various bands.

\subsection{The stellar distribution}
\label{stell.sec}


\begin{figure}
\centering 
\includegraphics[clip,bb=173 470 421 620]{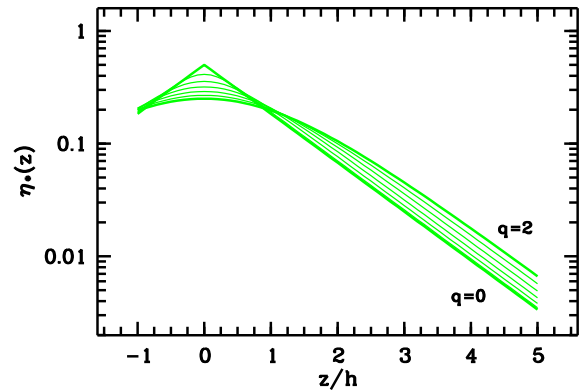}
\caption{The stellar emissivity $\eta_*(z)$ of the family of
$\sech^q$ models~(\ref{etasechq}). It is shown for different values of
the parameter $q$, ranging between the extreme values $q=0$ and $q=2$,
which are indicated and plotted in bold. It is clear that they all
have the same exponential asymptotic behaviour, whereas the sharpness
of the peak at $z=0$ is determined by $q$.}
\label{etasechq.ps}
\end{figure}


The vertical distribution of stars in disc galaxies is still a matter
of debate. The most straightforward way to derive such a vertical
distribution is to study the surface brightness of edge-on galaxies at
different heights above the plane. There is a general consensus that
at great heights the light distribution decreases exponentially. Close
to the plane of the galaxy however, dust attenuation makes the
observation of the vertical distribution difficult. As a consequence,
different models have been proposed. The most popular models are an
isothermal sheet distribution (van der Kruit \& Searle 1981) and an
exponential profile (Wainscoat et al.\ 1989). In order to allow a wide
range in vertical density distributions, we adopt a two-parameter
family of stellar emissivity profiles which were adapted from van der
Kruit~(1988),
\begin{equation}
	\eta_*(z) 
	=
	\frac{1}{q\,B(\tfrac{q}{2},\tfrac{1}{2})\,h}
	\sech^q\left(\frac{z}{qh}\right),  
	\label{etasechq}
\end{equation}
where the function $B(x,y)$ in the represents the Beta function
(Abramowitz \& Stegun~1972). The first factor is determined such that
the normalization condition~(\ref{normeta}) is satisfied. At great
heights above the plane of the galaxy, the light
distribution~(\ref{etasechq}) indeed decreases exponentially,
\begin{equation}
	\eta_*(z)
	\propto
	{\text{e}}^{-|z|/h}
	\qquad\text{for $|z|\gg h.$}
\end{equation}
The first parameter in~(\ref{etasechq}) is the exponential scaleheight
$h$. It is generally known that the velocity dispersion of different
stellar types in the Galaxy is correlated with the age of the stars
(e.g.\ Dehnen \& Binney~1998). Due to this dynamical heating of the
disc, the vertical scaleheight of the stars increases as a function of
age (Wainscoat et al.\ 1992). Because the younger stellar populations
are bluer than older ones, it can be expected that the scaleheight of
stars increases with wavelength. Xilouris et al.\ (1999) accurately
modelled several edge-on disc galaxies and found no significant trend
of the stellar scaleheight with wavelength (from the blue to the
near-infrared $K$ band). Therefore we adopt a constant stellar
scaleheight.

The second parameter in~(\ref{etasechq}), $q$, corresponds to the
$2/n$ used by van der Kruit~(1988), and takes values between 0 and
2. It determines the shape of the stellar density near the plane of
the galaxy, and hence is called the shape parameter. More precisely,
the peak of the emissivity at $z=0$ becomes sharper as $q$ decreases.
A number of emissivity profiles for various values of $q$ is
illustrated in Figure~{\ref{etasechq.ps}}. For the special cases
corresponding to the values $q=0$, 1 and 2, we find the exponential,
the sech- and the isothermal disc respectively,
\begin{gather}
	q=0 \qquad
	\eta_*(z)
	=
	\frac{1}{2h}\,{\text{e}}^{-|z|/h}
	\\
	q=1 \qquad
	\eta_*(z)
	=
	\frac{1}{\pi h}\sech\left(\frac{z}{h}\right)
	\\
	q=2 \qquad
	\eta_*(z)
	=
	\frac{1}{4h}\sech^2\left(\frac{z}{2h}\right).
\end{gather}
Although the $\sech^q$ profiles can be adopted for any value of $q$
between 0 and 2 (e.g.~de Grijs et al.\ 1997), it are these three
models which are widely used in the literature to model the vertical
structure in disc galaxies.  For example, Schwarzkopf \&
Dettmar~(2000) investigated the surface brightness of a set of 110
edge-on disc galaxies, and divided them into these three classes. They
found that the intermediate sech-profile fitted the data (optical as
well as near-infrared) best for 55\% of the galaxies, whereas the
exponential and isothermal profiles fitted 35\% and 10\% of the
galaxies.

This would suggest to choose $q=1$ as the typical value for our
template model. However, in Section~{\ref{geom.sec}} we will
demonstrate that the choice of $q$ has only a minor effect on the
attenuation curve, such that we are relatively free to choose $q$ to
our convenience. We choose an exponential profile $q=0$ because it has
some computational advantages (see Appendix~{\ref{expmodel.sec}}).

\subsection{The dust distribution}
\label{dust.sec}

Beside stars, we also need a dust component in the galaxy, which is
determined by the opacity $\kappa(z)$. Again, the actual vertical
distribution of dust in disc galaxies is difficult to
investigate. Direct information about the distribution of dust can be
obtained by spatially resolved far-infrared or submm observations,
where the dust emission is directly observed. During the last years,
several nearby spiral galaxies have been modelled at these wavelengths
using ISO and SCUBA data (Haas et al.\ 1998, Alton et al.\ 1998a,
1998b, Davies et al.\ 1999, Domingue et al.\ 1999, Trewhella et al.\
2000). For NGC\,891, Alton et al.\ (2000) showed that the far-infrared
emission in very good agreement with the dust distribution derived
from radiative transfer calculations (Xilouris et al.\ 1998), where an
exponential vertical dust profile was assumed.

It seems therefore obvious to assume that stars and dust have a
similar distribution in the galaxy, but we allow the scaleheights to
be different. We thus adopt a family of opacity functions
\begin{equation}
	\kappa(z) 
	=	
	\frac{\tau_0}{q\,B(\tfrac{q}{2},\tfrac{1}{2})\,\zeta h}
	\sech^q\left(\frac{z}{q \zeta h}\right),
\label{kappasechq}
\end{equation}
which satifies the normalization condition~(\ref{deftau0}). For our
template model we will thus adopt an exponential model for both stars
and dust. The quantity $\zeta$ represents the layering parameter,
i.e.\ the ratio of the scaleheights of dust and stars (Disney et al.\
1989).

Combining the emissivity~(\ref{etasechq}) and the
opacity~(\ref{kappasechq}) we can calculate the source function
$S_*(\tau)$ for our galaxy models.\footnote[1]{In
Appendix~{\ref{expmodel.sec}}, we explicitly calculate the source
function for the exponential model.}  It is fairly straightforward to
check that $S_*(\tau)$ will have the form
\begin{equation}
	S_*(\tau) 
	= 
	\frac{1}{\tau_0}
	s_*\left(\frac{\tau}{\tau_0},\zeta,q\right),
\label{formS}
\end{equation}
where $s_*(t,\zeta,q)$ is normalized as
\begin{equation}
	\int_0^1 s_*(t,\zeta,q)\,\txd t = 1.
\label{norms}
\end{equation}
The source function hence does not depend on either of the
scaleheights of the stellar or dust distribution, but only on their
ratio $\zeta$. Therefore the same is true for the attenuation
$A(\mu)$, which will thus depend on two geometry parameters, the shape
parameter $q$ and the layering parameter $\zeta$.

What is a representative value for the layering parameter~?  Normally,
the interstellar matter sinks down to the central plane of a galaxy
and forms an obscuring layer which is narrower than the stellar layer,
such that $\zeta<1$. This is observed in the Galaxy, where the
scaleheights of the (thin) stellar disc and the dust disc are
approximately 300\,pc and 100\,pc respectively. In edge-on galaxies
dust lanes suggest that also there the dust distribution is narrower
than the stellar one. Using detailed radiative transfer modelling for
seven such galaxies, Xilouris et al.\ (1999) find that the dust
scaleheight is about half that of the stars. For our template model we
set $\zeta=0.5$, but in Section~{\ref{geom.sec}} we will consider a
wide range in $\zeta$ in order to investigate the dependence of the
relative star-dust geometry on the attenuation.

The last parameter in our galaxy model is the total $V$ band optical
depth $\tau_V$. Again, a realistic value for the optical depth in disc
galaxies has been a subject of debate for a long time already. An
extensive discussion can be found in e.g.\ Xilouris et al.\ (1997) and
Kuchinski et al.\ (1998). The most widely supported idea is that
spiral galaxies are transparant or moderately opaque if they are seen
face-on, with a typical $V$ band face-on optical depth of round
unity. For our template model we adopt $\tau_V=1$, which corresponds
to an optical depth of 3 in the far UV and 0.5 in the $I$ band
(Table~{\ref{optprop.tab}).

\section{The influence of scattering}
\label{scatimpo.sec}

\subsection{Approximations of the RTE}

The RTE as written in~(\ref{tv2}) is a partial integro-differential
equation, containing a differentiation along the path and an
integration over the angle. The scattering term is the one that makes
the RTE difficult, because this term is responsible for the coupling
of the RTE along different paths: due to the integration over $\mu$ we
cannot solve the RTE for different paths consecutively, but we have to
solve it for all paths at the same time. Therefore it would be very
practical if we could somehow get around the last term.

\subsubsection{No scattering}

One way is to neglect the scattered emission alltogether, i.e.~to take
into account the photons that are removed from the beam, but neglect
the photons that are added to the beam due to scattering. This is the
same as assuming that all the interactions between dust grains and
photons are absorption events, and hence that there are no
scatterings. We will denote the resulting attenuation curve as the
{\em ns} attenuation.  Mathematically this translates into setting
$\omega=0$ (the albedo is the ratio of the scattering efficiency to
the total extinction efficiency), such that the last term
in~(\ref{tv2}) vanishes, which turns the RTE into an ordinary
differential equation
\begin{equation}
	\mu\,\frac{\partial I}{\partial\tau}(\tau,\mu)
	=
	I(\tau,\mu)
	-S_*(\tau).
\end{equation}
With the appropriate boundary conditions it can directly be solved,
and the {\em ns} attenuation becomes
\begin{equation}
	A^{\text{ns}}(\mu) 
	=
	-2.5\log
	\int_0^{\tau_0}
	S_*(\tau)\,
	\exp\left(-\frac{\tau}{\mu}\right)
	\txd\tau.
\label{Ans}
\end{equation}

A second way of getting around the scattering term is to neglect
scattering completely, both the scattering absorption and the
scattering emission. The photons that would normally be scattered out
of the beam, are thus assumed to continue on their path. For the RTE
this means that not only the last term vanishes (the photons added to
the beam due to scattering), but also the fraction of the first term
(the photons removed from the beam due to scattering).  In this case
the RTE becomes
\begin{equation}
	\mu\,\frac{\partial I}{\partial\tau}(\tau,\mu)
	=
	(1-\omega)I(\tau,\mu)
	-S_*(\tau).
\label{pures}
\end{equation}
This kind of interaction can also be described as completely {\em
forward scattering}, because also then a scattering event has no net
effect on a beam. The ARF for pure forward
scattering is $\Psi(\mu,\mu') = 2\delta(\mu-\mu')$, and inserting
this in~(\ref{tv2}) we find the same equation~(\ref{pures}). This
equation is readily solved, and the {\em fs} attenuation is
\begin{equation}
	A^{\text{fs}}(\mu)
	=
	-2.5\log
	\int_0^{\tau_0}
	S_*(\tau)\,
	\exp\left[-\frac{(1-\omega)\tau}{\mu}\right]
	\txd\tau.
\label{Afs}
\end{equation}
It is fairly straightforward to see that {\em fs} attenuation is in
fact equivalent to {\em ns} scattering, but with the total optical
depth replaced by an effective optical depth $\tau_{\text{eff}} =
(1-\omega)\tau_0$.  If we thus obtain an expression for the {\em ns}
attenuation, we immediately find the corresponding one for the {\em
fs} attenuation.

\subsubsection{(Approximate) isotropic scattering}

In radiative transfer problems, much attention has always been paid to
isotropic scattering. In some physical situations the assumption of
isotropy can be correct. For example, according to Bruzual et al.\
(1988) and Corradi et al.\ (1996), the asymmetry parameter is close to
zero in the near-infrared $K$ band, such that scattering is virtually
isotropic there.\footnote{Note however that this assumption is not
always followed, e.g.\ Gordon et al.\ (1997) find $g_K=0.43$.} Another
useful situation where the assumption of isotropy can be applied is a
medium where photons have a large probability to be scattered more
than once, because multiple scattering tends to wash out the
anisotropy effects. This principle has been applied by Vansevi\v{c}ius
et al.\ (1997), who solve the RTE in a disc galaxy using the iteration
method. For the first scattering they adopt the general anisotropic
scattering, for the next scattering events isotropic scattering is
assumed.

For isotropic scattering the distribution of the angles after a
scattering event is uniform, such that the ARF is simply
$\Psi(\mu,\mu')=1$. A proper treatment of isotropic scattering hence
does not change the integro-differential character of the RTE, and
still requires the solution of the RTE for all path together. There
are however two ways to approximate isotropic scattering such that the
integration in the scattering term disappears.


\begin{figure*}
\centering 
\includegraphics[clip,bb=67 466 527 684]{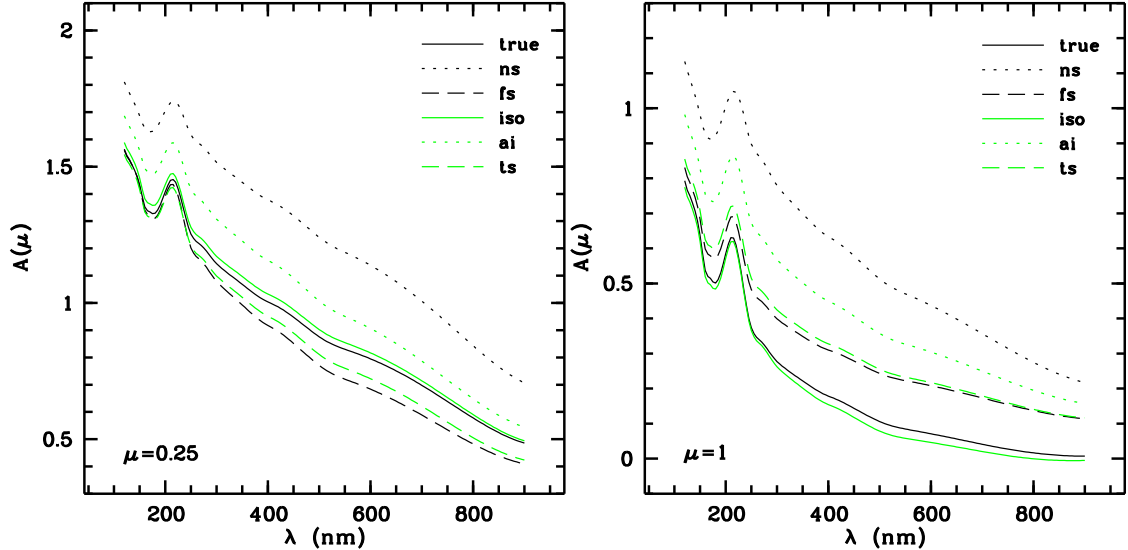}
\caption{The various attenuation curves $A(\mu)$ as a function of
wavelength. The left and right panel correspond to $\mu=0.25$ and
$\mu=1$ respectively. The adopted model is the template model.}
\label{scatimpo.ps}
\end{figure*}


\begin{figure}
\centering 
\includegraphics[clip,bb=173 466 423 684]{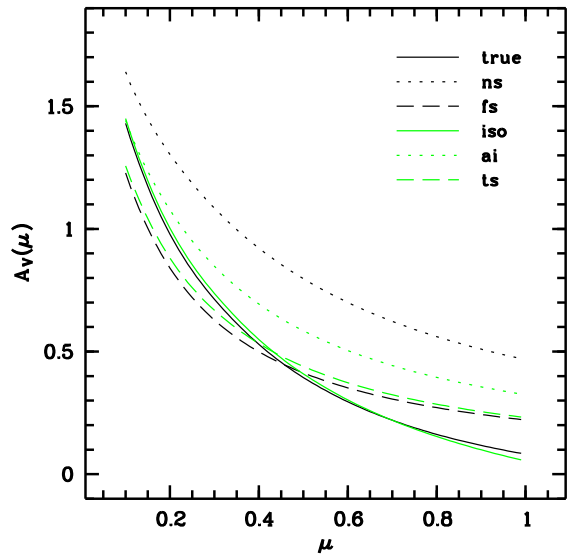}
\caption{The $V$ band attenuation curves $A(\mu)$ as a function of
$\mu$, corresponding to those in Figure~{\ref{scatimpo.ps}}.}
\label{scatimpo2.ps}
\end{figure}


The first possible approximation to isotropic scattering is the
so-called isotropic two-stream approximation, introduced by
Code~(1973)\footnote{Code~(1973) applied the two-stream approximation
to calculate the flux escaping from a spherical dust envelope
surrounding a star, and found a very good agreement with numerical
results. He also proposes a second approximate solution, which works
even better. However, this method cannot be applied to general
plane-parallel (and other) geometries in a straightforward way, and is
therefore not considered in this paper.}. In the {\em ts}
approximation a photon always remains on the same path, and after a
scattering event it will either move in the same direction (i.e.\
forward scattering), or it moves in the opposite direction (i.e.\
reflection). The probability of forward scattering and reflection are
equal. The ARF corresponding to this kind of scattering is thus
infinitely peaked in two directions, and can be written as
$\Psi(\mu,\mu') = \delta(\mu-\mu') + \delta(\mu+\mu')$.  The intensity
for a certain value of $\mu$ can then be found by solving the coupled
set of RTEs
\begin{gather}
	\mu\,\frac{\partial I}{\partial\tau}(\tau,\mu)
	=
	\left(1-\frac{\omega}{2}\right)I(\tau,\mu)
	-\frac{\omega}{2}I(\tau,-\mu)
	-S_*(\tau)
	\nonumber \\
	-\mu\,\frac{\partial I}{\partial\tau}(\tau,-\mu)
	=
	\left(1-\frac{\omega}{2}\right)I(\tau,-\mu)
	-\frac{\omega}{2}I(\tau,\mu)
	-S_*(\tau)
\end{gather}
For each inclination angle $\mu$ this set of ordinary differential
equations can be solved directly, such that we can calculate the
two-stream attenuation curve $A^{\text{ts}}(\mu)$.

Another approximation for isotropic attenuation has been introduced by
Natta \& Panagia (1984), and afterwards adopted by various authors
(e.g.\ Guiderdoni \& Rocca-Volmerange~1987, Calzetti et al.\
1994). Natta \& Panagia~(1984) suggest that a good approximation can
be achieved by considering an attenuation equivalent to {\em ns}
attenuation, but replacing the total optical depth by an effective
optical depth $\tau_{\text{eff}} = \sqrt{1-\omega}\,\tau_0$.  The
approximate isotropic attenuation or {\em ai} attenuation is thus
\begin{equation}
	A^{\text{ai}}(\mu) 
	=
	-2.5\log
	\int_0^{\tau_0}
	S_*(\tau)\,
	\exp\left(-\frac{\sqrt{1-\omega}\,\tau}{\mu}\right)
	\txd\tau.
\label{Aai}
\end{equation}
Although there is no physical principle behind this approximation, it is
attractive due to its simplicity.

\subsection{Comparison of the attenuation curves}

In the previous section we have described five approximations for the
attenuation curve $A(\mu)$. These are compared in the
Figures~{\ref{scatimpo.ps}} and~{\ref{scatimpo2.ps}}, where we plot
them respectively as a function of wavelength for two different angles
$\mu$, and as a function of angle for the $V$ band.

\subsubsection{Scattering versus no scattering}

We first compare the attenuation curve with the {\em ns}
approximation, i.e.\ the situation where the scattering emission is
neglected. Qualitatively, the difference between the {\em ns} and all
other attenuation curves is obvious: by not taking into account the
scattered emission, the attenuation is overestimated in all
directions. Because the probability of scattering is slightly higher
than the probability of absorption at optical wavelengths (see
Table~{\ref{optprop.tab}}), it will be no surprise that this
overestimation compared to the true attenuation can be
considerable. In particular for face-on directions, we obtain a
difference of nearly half a magnitude for our template model. Only for
high inclinations, the {\em ns} approximation is more or less
satisfactory.

Another way to understand the effect of scattering is to compare the
true attenuation curve with the {\em fs} attenuation curve, i.e.\ the
situation where scattering is neglected completely. Particularly in
Figure~{\ref{scatimpo2.ps}} the effect of including scattering on the
attenuation curve is clearly observable: the attenuation becomes
larger for small $\mu$ and smaller for large $\mu$. The overall effect
of scattering is thus apparently that photons are removed from lines
of sight with a high inclination and sent into face-on directions,
where they leave the galaxy. This is due to the fact that the optical
depth along a path with inclination $\mu$ is proportional to
$1/\mu$. Photons that initially (or after a scattering event) move on
a path with large inclination have a large possibility to interact
with a dust grain. If they are scattered into a direction which is
nearly face-on, the probability to interact with another dust grain is
much smaller, such that they can easily leave the galaxy.


\begin{figure}
\centering 
\includegraphics[clip,bb=173 466 423 684]{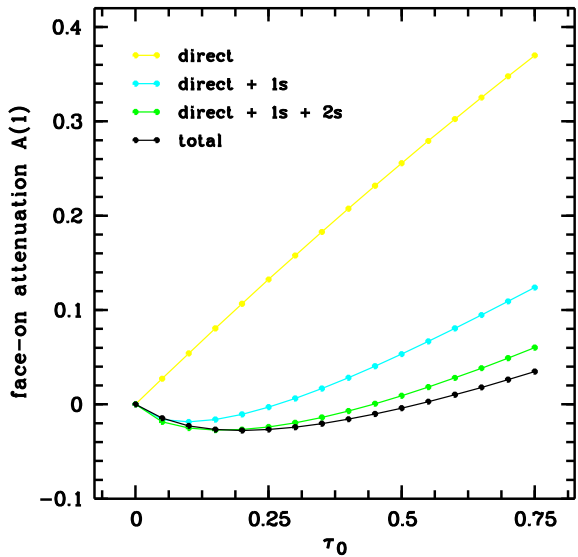}
\caption{The $V$ band face-on attenuation $A(1)$ as a function of the
total optical depth $\tau_0$. Shown are the attenuation corresponding
to photons that leave the galaxy directly, directly or after one
scattering (direct + 1s), directly or after one or two scatterings
(direct + 1s + 2s), and the total attenuation. Except for the varying
total optical depth, the galaxy models have the same parameters as our
template model ($q=0$ and $\zeta=0.5$.)}
\label{brighter.ps}
\end{figure}


This overall effect of scatterig is responsible for a rather strange
behaviour for face-on galaxies. If we look carefully at the right
panel in Figure~{\ref{scatimpo.ps}}, we see that the true face-on
attenuation becomes zero at the $I$ band. The loss of radiation due to
absorption ($0.2~{\text{mag}}\approx20$ per cent) is thus completely
balanced by the gain due to scattering. This is not due to some
particular properties of the dust at the $I$ band, but only to the
smaller optical depth. When we consider a smaller total optical depth,
the extinction can even be negative, i.e.\ a face-on galaxy can be
brighter than it would be without dust. This is illustrated in
Figure~{\ref{brighter.ps}}, where we plot the face-on attenuation in
the $V$-band as a function of the total optical depth. Using the
iteration method such that the contribution of each partial intensity
can be considered separately. The attenuation corresponding to the
photons that leave the galaxy directly, i.e.\ the {\em ns}
attenuation, is of course positive everywhere. But adding the photons
which leave the galaxy after one scattering event, the attenuation
already is negative for $\tau_0<0.25$. The total attenuation remains
negative for $\tau_0<0.5$, a result in correspondence with Bruzual et
al.\ (1988) and Di Bartolomeo et al.\ (1995).

\subsubsection{Anisotropic versus isotropic scattering}

At first sight one would expect a significant difference between the
isotropic and the anisotropic attenuation curves. First, at optical
and UV wavelengths the anisotropy parameter is fairly large
($g\sim0.5$), and for these wavelengths the Henyey-Greenstein phase
function is far from isotropic (e.g.\ Bianchi et al.\ 1996,
Figure~2). Second, the total optical depth in our galaxy model is only
moderate, such that the anisotropy effects should not be washed out by
multiple scattering. However, Figure~{\ref{scatimpo.ps}}
and~{\ref{scatimpo2.ps}} show that the isotropic attenuation
approximates the anisotropic attenuation very well, with differences
between both attenuation curves only a few hundredths of a magnitude.

In Figure~A1 of Paper~I we plotted the Henyey-Greenstein ARF for
various values of $\mu'$. For nearly face-on lines-of-sight $\mu'$,
the distribution of the new angles $\mu$ is far from uniform. For very
inclined lines-of-sight however, the ARF is fairly uniform, due to the
averaging out over all azimuths. For photons which are scattered from
such lines-of-sight, the scattering seems thus approximately
isotropic. In our galaxy model, the majority of the scattering events
occurs on paths which are nearly edge-on, because along such paths the
optical depth is much larger than on paths which are nearly
face-on. This explains why even for the moderate optical depth of our
galaxy, the differences between the isotropic and anisotropic
attenuation are fairly small.

As mentioned, there is little gain in the observation that the
isotropic attenuation curve approximates the true anisotropic one so
well, because the approximation does not simplify the RTE. However, if
the {\em ts} or {\em ai} attenuation curves were to be satisfying
approximations for the isotropic attenuation curve, the operational
gain would be significant. Unfortunately, the
Figures~{\ref{scatimpo.ps}} and~{\ref{scatimpo2.ps}} show that the
approximations are not satisfying. The two-stream approximation is
both qualitatively and quantitatively comparable to the {\em fs}
approximation, i.e.\ an overestimation of the attenuation in face-on
directions of more than 0.1~mag, and an underestimation of the same
order for highly inclined directions. The {\em ai} attenuation curve
lies between the {\em ns} and {\em fs} attenuation curve, which is
logical if we compare the expressions~(\ref{Ans}), (\ref{Afs})
and~(\ref{Aai}). This means that the approximation is satisfying for
highly inclined directions, but the attenuation is severily
overestimated in the face-on direction ($\Delta A\approx 0.25$~mag).

\section{The influence of geometry}
\label{geom.sec}


\begin{figure}
\centering \includegraphics[clip,bb=166 467 411
657]{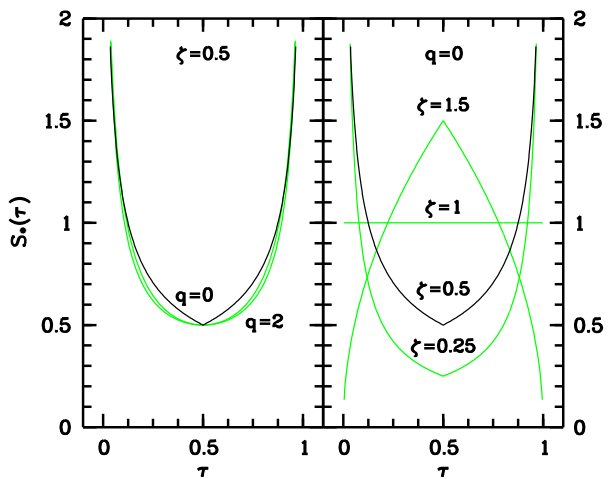}
\caption{The source function $S_*(\tau)$ of the family of $\sech^q$
models as a function of the parameters $q$ and $\zeta$. In each of the
panels the template model is the black curves, and we vary one of the
geometry parameters. {\em Left:}~variation of the shape parameter $q$,
taking the values 0, 1 and 2. {\em Right:}~variation of the layering
parameter $\zeta$, taking the values 0.25, 0.5, 1 and 1.5.}
\label{sourcesechq.ps}
\end{figure}


Our models contain two parameters which specifically determine the
geometry of the galaxy model: the shape parameter $q$ determines the
{\em actual distribution} of stars and dust near the galaxy plane,
whereas the layering parameter determines the {\em relative
distribution} of stars and dust. In this section we investigate to
which degree these geometry parameters determine the attenuation
curve of the galaxy.

\subsection{The shape parameter $q$}

The RTE~(\ref{tv2}) shows that the source function completely
determines the attenuation curve, i.e.\ once $S_*(\tau)$ is known, we
can calculate $A(\mu)$. Therefore, if we want to know the influence of
$q$ on the attenuation curve it is interesting to firstly study its
influence on the source function. In Figure~{\ref{sourcesechq.ps}} we
plot $S_*(\tau)$ for various models with different shape
parameters. The source function depends only weakly on the parameter
$q$, more precisely the shape of $S_*$ becomes slightly rounder at the
centre of the galaxy as $q$ increases. It seems logical to predict a
similar insensitivity on $q$ for the attenuation curves. This is
confirmed by our calculations. Even the differences between the
attenuation curves of the two extreme cases, the exponential disc and
the isothermal sheet, are almost negligibly small.

\subsection{The layering parameter $\zeta$}
\label{zeta.sec}


\begin{figure*}
\centering 
\includegraphics[clip,bb=54 435 540 657]{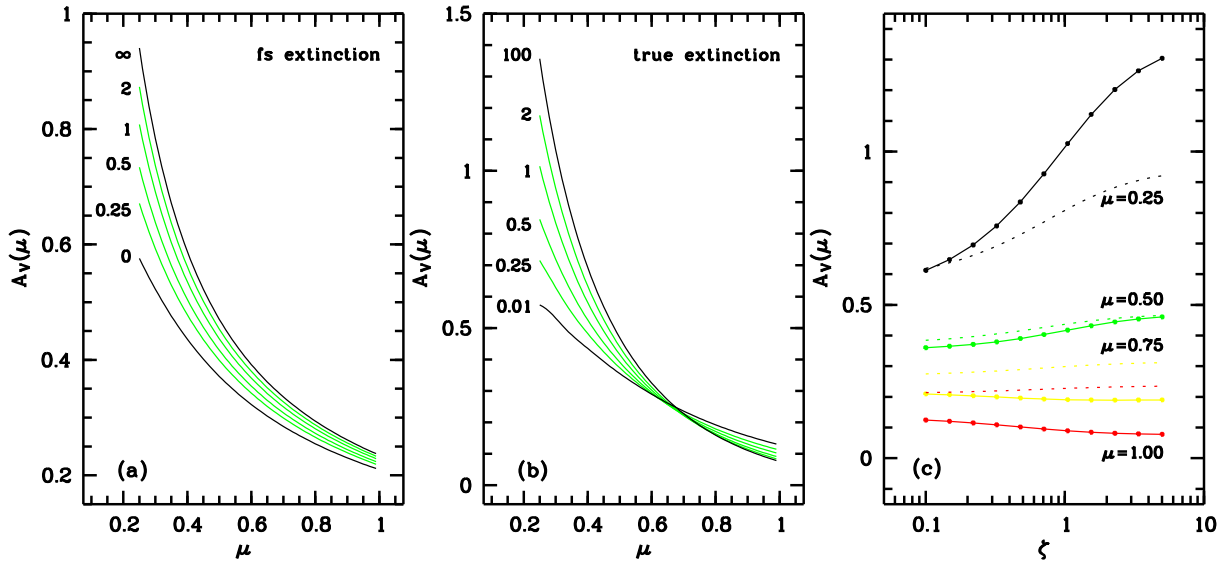}%
\caption{({\bf a.})~The {\em fs} attenuation $A_V(\mu)$ curve as a
function of $\mu$ for some different values of the layering parameter
$\zeta$, which are indicated. The black curves denoted by $\zeta=0$
and $\zeta=\infty$ correspond to the thin dust disc and overlying
sheet models respectively. ({\bf b.})~Same as a., but now the true
attenuation curve. ({\bf c.})~The true and {\em fs} attenuation curves
$A_V(\mu)$ as a function of the layering parameter $\zeta$. The solid
lines represent the true attenuation curve, the dotted lines are the
{\em fs} attenuation curves. The results are shown for four angles,
which are indicated.}
\label{depzeta.ps}
\end{figure*}


The layering parameter is the ratio of the scaleheights of dust and
stars and hence determines the relative distribution of the two
components. Variation of this parameter yields a wide range of
dust-stars geometries. This can be seen in
Figure~{\ref{sourcesechq.ps}}, where the source function is plotted
for various models with different layering parameter
$\zeta$. According to the shape of the source function for a
particular value of $\zeta$, we can classify the different geometries
into three classes.

For $\zeta=1$, the stars and dust have the same scaleheight and thus
the same spatial distribution. The source function is then independent
of the depth in the galaxy, and due to the
normalization~(\ref{normeta}) we find $S(\tau) =
1/\tau_0$. In general, each galaxy model with a similar
distribution of stars and dust will have such a constant source
function. Because the attenuation curve is completely determined by
the source function only, all galaxy models with a similar
distribution of stars and dust will be equivalent. In particular, our
model will be equivalent to the {\em homogeneous slab}, i.e.~a finite
galaxy model with stars and dust homogeneously mixed. Such a model has
been considered by many authors (e.g.\ Bruzual et al.\ 1988, Calzetti
et al.\ 1994, Di Bartolomeo et al.\ 1995) as an approximation of a
disc galaxy.

For $\zeta<1$ the dust has a smaller scaleheight than the stars, such
that the attenuation will predominantly occur in the central
regions. As mentioned in Section~{\ref{dust.sec}}, this is the most
realistic range for the layering parameter. If $\zeta<1$, the source
function has its minimal value in the centre of the galaxy, and
diverges at the edges. If $\zeta$ tends to zero, the dust will form an
infinitely thin obscuring layer in the central plane of the galaxy. We
will refer to this model as the {\em thin dust disc} model.

For $\zeta>1$ the dust has a larger scaleheight than the stars, and
the attenuation will thus occur relatively more in the outer regions
of the galaxy. The source function is then maximal in the centre of
the galaxy, and vanishes at the edges of the galaxy. If $\zeta$ grows
larger, the bulk of the dust attenuation will occur further away from
the central plane of the galaxy. In the limit $\zeta\rightarrow\infty$
the dust is infinitely thinly distributed over all space, and will
effectively behave as a set of two obscuring layers on both sides of
the galaxy. This geometry, where the obscuring material is located
only between the observer and the source, is known as the {\em
overlying screen} model. It is the analogue of attenuation of stars in
the Galaxy. Various authors advise against the use of this model
(Disney et al.\ 1989, Witt et al.\ 1992).

The source function is thus strongly dependent on the relative
distribution of stars and dust, whereas the actual shape of the
distribution is only of minor importance. We will now investigate in
detail is whether this also means that the influence of the layering
parameter on the attenuation curve is significant.

\subsubsection{Effects on the {\em fs} attenuation curve}

First we will only consider true absorption and neglect scattering,
and we will hence investigate the influence of $\zeta$ on the {\em fs}
attenuation curve. For the exponential model $q=0$, the source
function is fairly simple, and the integration~(\ref{Afs}) can be
performed analytically. One finds
\begin{equation}
	A^{\text{fs}}(\mu)
	=
	0.543\,\frac{(1-\omega)\tau_0}{\mu}
	-2.5\log
	{\cal W}_{\zeta}\left[\frac{(1-\omega)\tau_0}{2\mu}\right].
\label{Azeta}
\end{equation}
This expression is derived in Appendix~{\ref{expmodel.sec}}, and the
definition and some properties of the function ${\cal W}_\zeta(x)$ are
described in Appendix~{\ref{wzeta.sec}}. The advantage of the
expression~(\ref{Azeta}) is that it easily allows us to study the
attenuation curve as a function of $\zeta$. For every value of $x$,
${\cal W}_\zeta(x)$ is a decreasing function of $\zeta$, moderately
decreasing for small $x$ and strongly decreasing for large $x$. This
implies that $A^{\text{fs}}(\mu)$ will be a increasing function of
$\zeta$. The more extended the dust distribution, relative to the
stellar one, the larger the attenuation and thus the fainter the
galaxy appears. In particular, the overlying screen model is the most
efficient and the thin dust disc the least efficient absorbing
geometries. This behaviour is illustrated in
Figure~{\ref{depzeta.ps}a}, where we plot the {\em fs} attenuation
curve $A^{\text{fs}}(\mu)$ as a function of $\mu$ for a set of
layering parameters $\zeta$. For highly-inclined directions (small
values of $\mu$) the {\em fs} attenuation changes considerably as a
function of the layering parameter.

\subsubsection{Effects on the true attenuation curve}

Let us now investigate the influence of $\zeta$ on the true
attenuation curve, which has to be calculated numerically. In
Figure~{\ref{depzeta.ps}b} the results are shown: we see a different
behaviour between the {\em fs} and the true attenuation
curves. Whereas the {\em fs} attenuation is a increasing function of
$\zeta$ in all directions, the true attenuation is not: for small
inclination angles, the true attenuation decreases as a function of
$\zeta$. This difference is also illustrated in
Figure~{\ref{depzeta.ps}c}, where we plot the {\em fs} and true
attenuation explicitly as a function of the layering parameter for
different inclination angles. The {\em fs} curves all rise, whereas
the true attenuation curve rises for small values of $\mu$, but
decreases in the face-on direction $\mu=1$.

If scattering is not taken into account, galaxies with an extended
dust distribution are more efficiently obscured than galaxies with
their dust concentrated in the central region. However, if scattering
is taken into account, the efficiency turns around for face-on
directions. {\em In face-on directions, galaxies with their dust
concentrated in the central region are more efficiently obscured than
galaxies with an extended dust distribution}. How can this be
understood physically? In Section~{\ref{scatimpo.sec}} we showed that
scattering biased the preferential direction for photons to leave the
galaxy: it is easier to leave it face-on than for highly-inclined
directions. In particular photons which are scattered near the edge of
the galaxy are responsible for this: if a photon is scattered there
into a face-on direction, its probability to leave the galaxy without
another interaction is high. Now, the more extended the dust
distribution in the galaxy, the more dust in the outer regions, and
the more photons which are scattered out of the galaxy in nearly
face-on directions. Relative to galaxies with concentrated dust,
galaxies with an extended dust distribution on the one hand
efficiently absorbe photons and on the other hand they scatter a
considerable amount of photons in the face-on direction.


\begin{figure}
\centering 
\includegraphics[clip,bb=173 313 421 620]{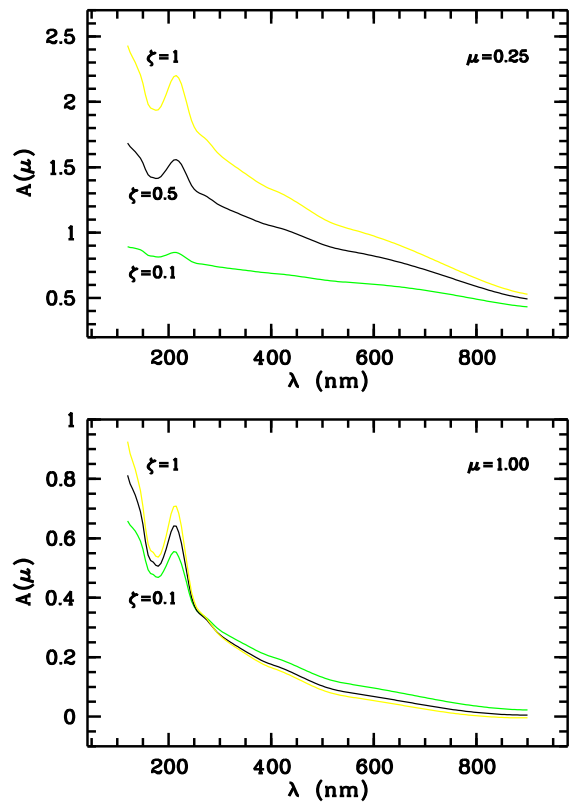}
\caption{Three attenuation curves $A(\mu)$ as a function of
wavelength for three models with a different layering parameter. The
values for $\zeta$ are 0.1, 0.5 and 1, the other parameters are those
of the template model ($\tau_V=1$ and $q=0$). The upper panel and
the lower panel correspond to $\mu=0.25$ and $\mu=1$ respectively.}
\label{shape.ps}
\end{figure}


Which of these two effect dominates is determined by the total optical
depth. For small optical depths the scattering effect will be more
important, and $A(\mu)$ will decrease with increasing $\zeta$ for
nearly-face-on directions. In contrary, for large optical depths, the
absorption will dominate and $A(\mu)$ will be an increasing function
of $\zeta$. We can thus conclude that
\begin{itemize}
\item inclined galaxies with a moderate optical depth are most
efficiently obscured if the dust is more extended than the stars.
\item nearly face-on galaxies with a moderate optical depth are most
efficiently obscured if the dust is concentrated in the central
regions.
\item optically thick galaxies are most efficiently obscured if the
dust is more extended than the stars.
\end{itemize}
Notice that the difference between small and large optical depth can
also be considered for one single galaxy, when we study the
attenuation as a function of wavelength. This is shown in
Figure~{\ref{shape.ps}}, where we plot the attenuation curve as a
function of wavelength for two angles $\mu$ and for three models with
a different layering parameter. For a total $V$ band optical depth
$\tau_V=1$, the attenuation behaves as an optically thick galaxy at
the blue side of the spectrum, and as an optically thin galaxy in the
red wavelength range, with the crossover between the two regimes
occurs around 300~nm ($\tau_0\approx2$).

These results can be compared with those found by Bruzual et al.\
(1988), who adopted a sandwich model (i.e.~a slab where stars and dust
are homogeneously mixed, surrounded on both sides by a slab containing
stars only). They varied the relative thicknesses of the slabs, and
found a similar result of the dependence of the attenuation curve on
the layering parameter (only shown for the face-on direction
$\mu=1$). Bruzual et al.\ (1988) concluded that the differences in the
attenuation curve between the various models have the same order of
magnitude than the uncertainties due to an imprecise knowledge of the
optical parameters of the dust, and hence that a homogeneous slab is a
satisfying model for disc galaxies. We agree that in the face-on
direction, quantitatively the difference in attenuation between
centrally concentrated and extended distributions is of the order
$\Delta A\approx0.1$~mag for optical wavelengths. These differences
become even smaller if we constrain the range of $\zeta$ to realistic
values. The layering parameters obtained by Xilouris et al.~(1999) all
lie between 0.30 and 0.75 at optical wavelengths, and for these values
$\Delta A(\mu)$ is only of the order of a few hundredths mag. Also for
moderately inclined galaxies, the attenuation curve will not seriously
depend on $\zeta$ if it this parameter is constrained to a realistic
range. Only if the optical depth is considerably higher, due to (a
combination of) a larger amount of dust, a more opaque wavelength
region or a large inclination, the uncertainty of the relative
geometry of dust and stars causes a considerable uncertainty on the
attenuation.

\section{Discussion and conclusion}

The aim of this paper was to investigate how the RTE can be simplified
and what errors are introduced by doing so. More than a decade ago
already, Bruzual et al.\ (1988) investigated to the effects of
scattering on the attenuation in disc galaxies (albeit for a simple
plane-parallel model), and Disney et al.\ (1989) showed that the
adopted geometry of the galaxy model strongly influences the
attenuation (without taking scattering into account however). We
wanted to reexamine these effects in detail and in the same manner,
i.e.\ within a realistic set of disc galaxy models and properly taking
multiple scattering into account.

In Section~{\ref{scatimpo.sec}} we showed that the nature of the
physical process of scattering is such that it connects different
lines of sight with each other. The overall effect of scattering is
that many photons which originally move on a inclined line-of-sight,
leave the galaxy in the face-on direction. Any approximate method to
simplify the RTE where photons cannot leave their initial path (such
as the {\em ns}, {\em fs}, {\em ts} and {\em ai} approximations
described in Section~{\ref{scatimpo.sec}}), will not be able to
reproduce this behaviour. In particular for face-on galaxies, the
optical depth will always be underestimated by such approximations, as
illustrated in Figure~{\ref{scatimpo.ps}}. If we compare the
attenuation curves corresponding to isotropic and anisotropic
scattering however, we see that the difference between them are only
of the order of a hudredth of a magnitude. Anisotropy effects are
essentially washed out. Inversely it will be impossible to determine
an ARF from photometry.

The effects of scattering are also important if we investigate the
influence of the star-dust geometry on the attenuation in disc
galaxies. If scattering is not taken into account, as in Disney et
al.\ (1989), we find that the attenuation is more effective if the
dust distribution is extended. For a same amount of dust, the larger
the layering parameter, the larger the attenuation. In particular this
means that the overlying screen geometry is the most efficient
obscuring geometry, whereas galaxies with a dust scaleheight smaller
than the stellar scaleheight, are less efficiently obscured. However,
when scattering is taken into account, this picture does not hold
anymore. We find the remarkable effect of scattering that in the
face-on direction, galaxies with a moderate optical depth ($\tau_0<2$)
are most efficiently obscured by a centrally concentrated dust
distribution. This is very interesting because the optical depth of
spiral galaxies is thought to be of the order unity in optical bands.

The effects of scattering and geometry not only need to be
investigated in a qualitative way, it is important to know them
quantitatively. This allows us to answer questions as how large our
error is when we approximate anisotropic scattering by assuming it is
forward, or when we assume a dust scaleheight equal to that of the
stars, where in reality it is only half of it.

The magnitude of these errors depends on the inclination. In the
face-on direction, the errors induced by not properly taking
scattering into account are largest. In the {\em ns} approximation,
the induced errors rise up to nearly half a magnitude, for forward
scattering they are of the order of 0.15~mag at optical wavelengths
(Figure~{\ref{scatimpo.ps}}b). The errors introduced by uncertainty
about the star-dust geometry are only of the order of a few hundredths
of a magnitude (Figure~{\ref{shape.ps}b}), and thus significantly
smaller. For inclined directions, the optical depth along
lines-of-sight towards the observer is much larger. The attenuation
curve then becomes less dependent on the angular redistribution
function, in agreement with Di Bartolomeo et al.\ (1995). The
differences between the attenuation curves corresponding to different
star-dust geometries grows significantly. For example, for a galaxy
with an inclination of 75$^\circ$ ($\mu=0.25$), the differences
between the true, isotropic, forward and two-stream attenuation curves
become about 0.1~mag at optical wavelengths, and only a few hundredth
of a magnitude in the UV (Figure~{\ref{scatimpo.ps}a). However,
assuming a homogeneous slab model instead of our template model, the
induced error on the attenuation curve is about 0.3~magnitudes at
optical and more than half a magnitude at UV wavelengths. For face-on
galaxies it is thus important to include scattering in a proper way in
the modelling, for inclined galaxies it is important to have a fairly
good idea of the star-dust geometry of the galaxy.

In the light of the adopted models, two remarks are
appropriate. (1)~By working in plane-parallel geometry we have the
advantage that we could use four independent methods to solve the RTE,
such that the accuracy of the obtained results can be checked
carefully. We are well aware of the fact that plane-parallel galaxy
models do not represent very realistic galactic discs. More detailed
modelling requires an exponential fall-off in the radial direction
(Freeman~1970), a distinction in attenuation between arm and interarm
regions (e.g.\ White et al.\ 2000), and an inclusion of the effects of
clumping of both stars and dust (Witt \& Gordon 1996, 2000; Bianchi et
al.\ 2000). Nevertheless we are convinced that our modelling can
contribute to the understanding of the mechanism of radiative
transfer, and we have all reasons to believe that our results would
hold qualitatively in more complex models. (2)~We know that scattering
and geometry are not the only two elements that introduce
uncertainties to the attenuation of disc galaxies. In Article~I we
solved the RTE for identical galaxy models, but with two different
sets of optical dust parameters found in the literature. We obtained
differences between the attenuation curves of the order of a few
tenths of a magnitude. Uncertainty about the optical properties of the
dust hence also introduces errors to the attenuation curve. However,
these errors are independent of those introduced by an inaccurate
treatment of scattering or geometry, and they hence do not affect the
conclusion of this work.

As a final remark, we re-iterate a warning that has also been issued
by others: do not use too simplistic models to determine the opacity
in disc galaxies. There are many ways to investigate the amount of
attenuation or the optical thickness of disc galaxies. Following
Xilouris et al.\ (1997), we can divide them into two categories, the
small $N$ approach and the large $N$ approach. The small $N$ approach
tries to determine the opacity of individual nearby disc galaxies by
constructing detailed radiative transfer models. Obviously, these
methods would reveal wrong results if scattering is neglected. The
large $N$ approach aims at determining a mean value for the optical
depth in disc galaxies, by statistically studying a large sample of
galaxies. Of such a sample of galaxies a certain observable is then
compared to a galaxy model, in order to derive the optical
depth. However, these methods are difficult to interpret because of
selection effects (Davies et al.\ 1993), and there are often
model-dependent. For example, Saunders et al.\ (1990) estimate an
average effective optical depth of disc galaxies by fitting a screen
model to the optical and far-infrared luminosity functions. Trewhella
et al.\ (1997) redid these calculations with a sandwich model, and
found a very different conclusion. We want to stress that also in such
studies the effects of scattering need to be taken into account,
because they are, except for very inclined galaxies, at least as
important as the effects of geometry. An often used prescript to take
scattering into account, is to determine an effective optical depth
$\tau_{\text{eff}}$ for a galaxy model without scattering and
compensate the obtained value for scattering. These compensations are
simply multiplications, such as $\tau_0=\tau_{\text{eff}}/(1-\omega)$
or $\tau_0=\tau_{\text{eff}}/\sqrt{1-\omega}$, which can but not
necessarily need to have a physical justification. With our paper we
hope to have clearly showed that such approximations are not able to
accurately reproduce the effects of scattering. We therefore argue
that scattering needs to be accounted for properly in statistical
studies concerning the opacity of disc galaxies.

\appendix

\section{The exponential model}
\label{expmodel.sec}

In this appendix we take a closer look at the exponential model $q=0$,
characterized by
\begin{gather}
	\eta_*(z) 
	=
	\frac{1}{2h}
	{\text{e}}^{-|z|/h}
	\label{expeta}
	\\
	\kappa(z) 
	=
	\frac{\tau_0}{2\zeta h}
	{\text{e}}^{-|z|/\zeta h}.
	\label{expkappa}
\end{gather}
Disney et al.\ (1989) already considered a similar galaxy model (their
Triplex model). They managed to calculate an expression for the
observed intensity if scattering is not taken into account. We will
derive the solution of the RTE by direct integration of the formal
solution~(\ref{Ans}). Therefore we first need an explicit expression
for the source function $S_*(\tau)$. The optical depth is obtained by
combining~(\ref{expkappa}) with~(\ref{optdepth}),
\begin{equation}
	\tau(z)
	=
	\begin{cases}
	\,\,\displaystyle
	\frac{\tau_0}{2}\left[
	2-\exp\left(\frac{z}{\zeta h}\right)\right]
	&
	\qquad\text{for $z\leq0$}
	\\[3mm]
	\,\,\displaystyle
	\frac{\tau_0}{2}
	\exp\left(-\frac{z}{\zeta h}\right)
	&
	\qquad\text{for $z\geq0$,}
	\end{cases}
\end{equation}
and inverting this relation we obtain
\begin{equation}
	z(\tau) 
	=
	\begin{cases}
	\,\,\displaystyle
	-\zeta h\ln(2t)
	&
	\qquad\text{for $0\leq t\leq\tfrac{1}{2}$}
	\\
	\,\,\displaystyle
	-\zeta h\ln\bigl[2(1-t)\bigr]
	&
	\qquad\text{for $\tfrac{1}{2}\leq t\leq1$,}
	\end{cases}
\end{equation}
where $t=\tau/\tau_0$ is the relative optical
depth. Introducing this expression in~(\ref{expeta})
and~(\ref{expkappa}) we find an expression for the stellar source
function $S_*(\tau)$,
\begin{equation}
	S_*(\tau)
	=
	\begin{cases}
	\,\,\dfrac{\zeta}{\tau_0}(2t)^{\zeta-1}
	& \text{for $0\leq t\leq\tfrac{1}{2}$}
	\\[3mm]	
	\,\,\dfrac{\zeta}{\tau_0}\bigl[2(1-t)\bigr]^{\zeta-1}
	& \text{for $\tfrac{1}{2}\leq t \leq 1$.}
	\end{cases}
\end{equation}
The source function is of the form~(\ref{formS}), and the
normalization condition~(\ref{norms}) is satisfied. The fact that
$S_*(\tau)$ has a fairly simple power law form, will allow us to
calculate an expression for the {\em ns} attenuation, by direct
integration of the formal solution~(\ref{Ans}). Inserting $S_*(\tau)$
in~(\ref{Ans}) we obtain
\begin{multline}
	A^{\text{ns}}(\mu)
	=
	0.543\frac{\tau_0}{\mu}
	\\
	-2.5\log\zeta\int_0^1
	x^{\zeta-1}\cosh\left[\frac{\tau_0}{2\mu}(x-1)\right]\txd x,
\end{multline}
which becomes, using the function ${\cal W}_\zeta(x)$ defined as in
Appendix~B,
\begin{equation}
	A^{\text{ns}}(\mu)
	=
	0.543\,\frac{\tau_0}{\mu}
	-2.5\log
	{\cal W}_{\zeta}\left(\frac{\tau_0}{2\mu}\right),
	\label{Aexp2}
\end{equation}
This result is in agreement with the solution obtained by Disney et
al.\ (1989). Knowing this expression, we immediately also have an
expression for the {\em fs} and {\em ai} attenuation curves
\begin{gather}
	A^{\text{fs}}(\mu)
	=
	0.543\,\frac{(1-\omega)\tau_0}{\mu}
	-2.5\log
	{\cal W}_{\zeta}\left[\frac{(1-\omega)\tau_0}{2\mu}\right]
	\\
	A^{\text{ai}}(\mu)
	=
	0.543\,\frac{\sqrt{1-\omega}\,\tau_0}{\mu}
	-2.5\log
	{\cal W}_{\zeta}\left(\frac{\sqrt{1-\omega}\,\tau_0}{2\mu}\right).
\end{gather}

\section{The function ${\cal W}_\zeta(\lowercase{x})$}
\label{wzeta.sec}


\begin{figure}
\centering 
\includegraphics[clip,bb=190 468 404 682]{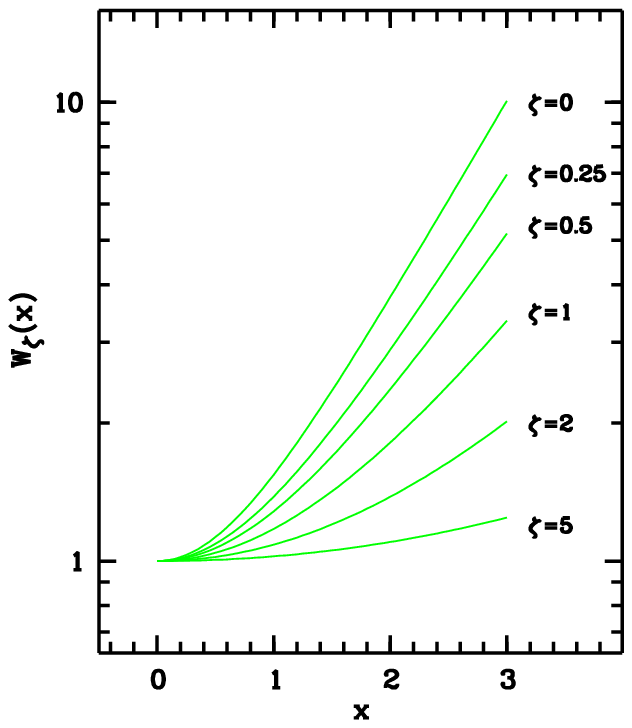}
\caption{The function ${\cal W}_\zeta(x)$ as a function of $x$ for
different values of $\zeta$.}
\label{wzeta.ps}
\end{figure}


For $\zeta>0$ and for all values of $x$, we define the function
${\cal{W}}_\zeta(x)$ by the integral
\begin{equation}
	{\cal{W}}_\zeta(x)
	=
	\zeta\int_0^1(1-t)^{\zeta-1}\cosh(xt)\,\txd t.
\end{equation}
For natural values of $\zeta$, the integral can immediately be solved
in terms of elementary functions, for example
\begin{gather}
	{\cal{W}}_1(x)
	=
	\frac{\sinh x}{x}
	\\
	{\cal{W}}_2(x)
	=
	2\,\frac{\cosh x-1}{x^2}
	\\
	{\cal{W}}_3(x)
	=
	6\,\frac{\sinh x+x}{x^3}.
\end{gather}
For non-integer values of $\zeta$ the integral can be calculated by
expanding the cosh function,
\begin{align}
	{\cal{W}}_\zeta(x)
	&=
	\zeta \sum_{k=0}^\infty	
	\frac{1}{(2k)!}\,x^{2k}\int_0^1(1-t)^{\zeta-1}\,t^{2k}\,\txd t
	\nonumber \\
	&=
	\sum_{k=0}^{\infty}
	\frac{\Gamma(\zeta+1)}{\Gamma(\zeta+2k+1)}\,x^{2k}
	\label{app_defW}
	\\
	&=
	1 + \frac{x^2}{(\zeta+1)(\zeta+2)} 
	\nonumber\\
	&\qquad\quad+ \frac{x^4}{(\zeta+1)(\zeta+2)(\zeta+3)(\zeta+4)}
	+ \cdots.
	\label{app_defW2}
\end{align}
Using d'Alembert's convergence theorem one can easily show that this
power series converges absolutely. It can be written as a
hypergeometric function,
\begin{equation}
	{\cal{W}}_\zeta(x)
	=
	{}_1\!F_2\left(1;\frac{\zeta+1}{2},
	\frac{\zeta+2}{2};\frac{x^2}{4}\right).
\end{equation}
Considering the expansion~(\ref{app_defW}) we can also consider the
special cases $\zeta=0$ and $\zeta=\infty$,
\begin{gather}
\label{app_wzetaspec}
	{\cal{W}}_0(x) = \cosh x \\
	{\cal{W}}_\infty(x) = 1.
\end{gather}
In Figure~{\ref{wzeta.ps}} we plot ${\cal{W}}_\zeta(x)$ as a function
of $x$ for a few values of $\zeta$. It is an even function of $x$,
increasing for positive $x$, the minimal value being
${\cal{W}}_\zeta(0)=1$ for all $\zeta$. Regarded as a function of
$\zeta$ for fixed values of $x$, ${\cal{W}}_\zeta(x)$ it is a
decreasing function, depending weakly on $\zeta$ for small $x$ and
strongly for large $x$.

\bsp
\end{document}